\documentstyle[editedvolume,psfig]{crckapb} 

\newcommand{\stt}{\small\tt}

\begin{opening}
\title{The Formation and Evolution of Galaxies:}
\subtitle{Perspectives on the Origin of the Hubble Sequence}
\author{R.S. ELLIS}
\institute{Institute of Astronomy\\
           Cambridge, UK}

\end{opening}

\runningtitle{FORMATION AND EVOLUTION OF GALAXIES}

\begin{document}

\begin{abstract}
I review recent observational progress concerning
the evolution of the morphological distribution of galaxies in the
rich cluster environment and in the faint field population.
By coupling HST imagery with ground-based spectroscopic
diagnostics, evidence accumulates that galaxy morphology
can be a transient phenomenon reflecting various changes in the
star formation rate. Possible physical processes responsible
for these changes are discussed. Future progress
in understanding them will depend on securing 2-D spectroscopic 
data for representative systems.\footnote{\stt To appear in IAU 
Symposium 183: Cosmological Parameters \& Evolution of the Universe, 
ed Sato, K.}
\end{abstract}

\section{Introduction}

It is a time of great progress in cosmology and galaxies remain the
most useful fabric for studying the Universe, both at the present and
early times. In this summary of a rapidly moving area, I have
restricted myself to three broad results with two connecting themes.
The first theme is concerned with the synergy achieved between space
and ground-based facilities, particularly that between the refurbished
Hubble Space Telescope (HST) and the new generation of 8-10 metre
telescopes. The second more fundamental theme is related to the
view that galaxy morphology is intimately connected with
the nature and history of star formation. The growing body of HST data
indicates that morphology may be a transient feature, reflecting short
timescale changes in the star formation rate. Whilst this complicates
traditional approaches towards galaxy evolution based on isolating
population subsets according to morphological classifications, it
provides an important insight into the physical processes governing star
formation on galactic scales.

Thus far, the greatest progress has been made in studies based on
properties integrated across the luminosity function i.e. for the
detected population. In quantative detail, these results remains
uncertain but new approaches based on the resolved properties of
distant galaxies, taking advantage of both aspects stressed above, will
allow us to study the physical processes responsible for the trends
seen in the integrated populations and hence assist in explaining the
distribution of morphological types seen at different epochs.

\section{The Role of the Environment}

One of the most fundamental observations relating to the origin of the
Hubble sequence is the morphology-density relation (Dressler 1980).
Does this dependence indicate that galaxy morphology was determined at
birth with subsequent evolution occurring in a `closed box'
approximation? Or does it imply that galaxies have been transformed
according to external processes? The first indications that the range
of galaxy types is formed, at least for some systems, by {\it nurture}
rather than {\it nature}, arose from the results of Butcher \& Oemler
(1978). Using the fraction of luminous cluster members whose rest-frame
colours were bluer than a fiducial value, those authors found the mean
star formation rate in the cores of rich clusters was significantly
higher 3-4 Gyr ago. HST images from a comprehensive survey of 10
clusters with 0.37$<z<$0.55 undertaken by the `Morphs' collaboration
(Smail et al 1997a, Ellis et al 1997, Dressler et al 1997) have
confirmed how this modest but important change in mean star formation
rate in cluster environments is accompanied by a radical shift in the
galaxy morphologies.  The bulk of the blue light originally located by
Butcher \& Oemler arises in disk galaxies (Figure 1) noticeably absent
in present day cluster cores.  Although Allington-Smith et al (1994)
showed, on the basis of ground-based multi-colour photometry, that the
rising blue fraction with redshift was a trend largely confined to rich
clusters, the Morphs team has provided the first glimpse of the
morphology-density relation at a redshift $\simeq$0.5 using their
extensive HST data.

\begin{figure}
\psdraft
\psfull
\caption{\it The transition in morphologies in the rich cluster
environment since a redshift $z\simeq$0.3. (Top) $R$ band ground-based
image of the central 200 $h^{-1}$ kpc of the Coma cluster (courtesy of
David Carter), (Bottom) HST F814W images of a comparable physical region in
the distant cluster AC118 ($z$=0.31)}.
\end{figure}

Before discussing the morphological distributions quantitatively, one
might question whether such classifications can be done reliably at
such limits, even noting the superb performance of HST. Morphology is,
by its nature, a somewhat subjective quantity (Naim et al 1995), but the
availability of HST data has spurred much activity, both in reducing
the subjective nature of galaxy classifications, in terms of indices
based on image concentration and asymmetry (Abraham et al 1994), and in
simulations which aim to quantify the effect of redshift on the visual
appearance of galaxies of known morphological type (Abraham et al
1996a). These studies indicate only modest shifts towards apparently
later types at $z>$0.7 arising from surface brightness and bandshifting
effects (Brinchmann et al 1997).

The Morphs data reveals a substantial change in the morphology-density
relation at high redshift (Figure 2). The abundance of spirals is
higher and less dependent on the projected surface density
notwithstanding the reduced range at high $z$ c.f. locally (1.5 dex
c.f. 2.5). Moreover, the {\it abundance} of S0s is 2-3 times less in
proportion at high redshift. Whilst S0s could clearly be preferentially
misclassified as Es, one would imagine this to be of greatest concern for
face-on examples. In fact, the population ellipticity distribution
shows no sign of a loss of face-on examples when compared to that
determined locally (Dressler et al 1997). Figure 2 suggests that 
present-day S0s must be counterparts of the intermediate star-forming 
disk galaxies and that the bulk of the Es are largely unaffected by 
whatever environmental processes occurred to the spirals.

\begin{figure}
\centerline{\psfig{file=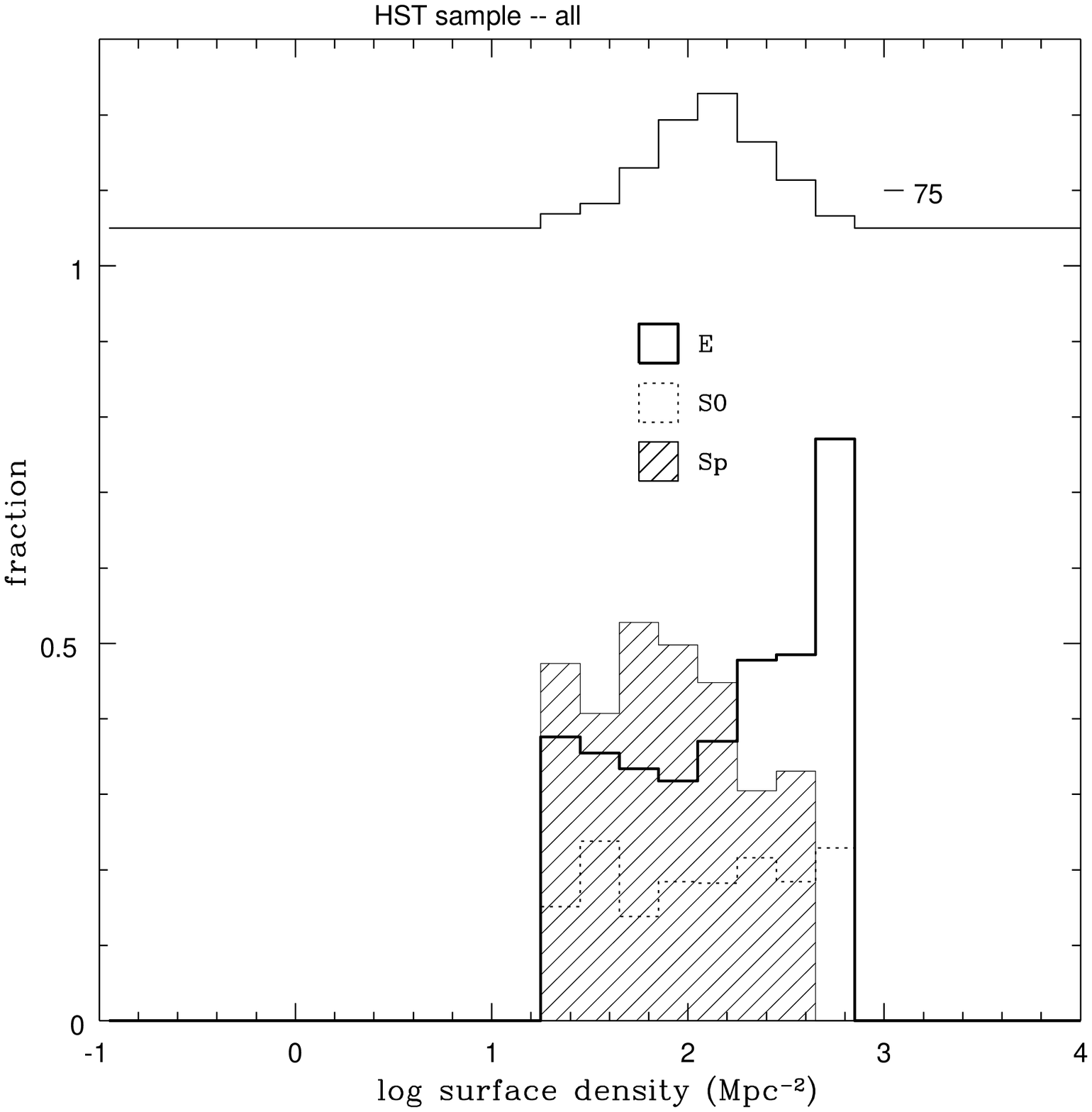,width=70mm}}
\centerline{\psfig{file=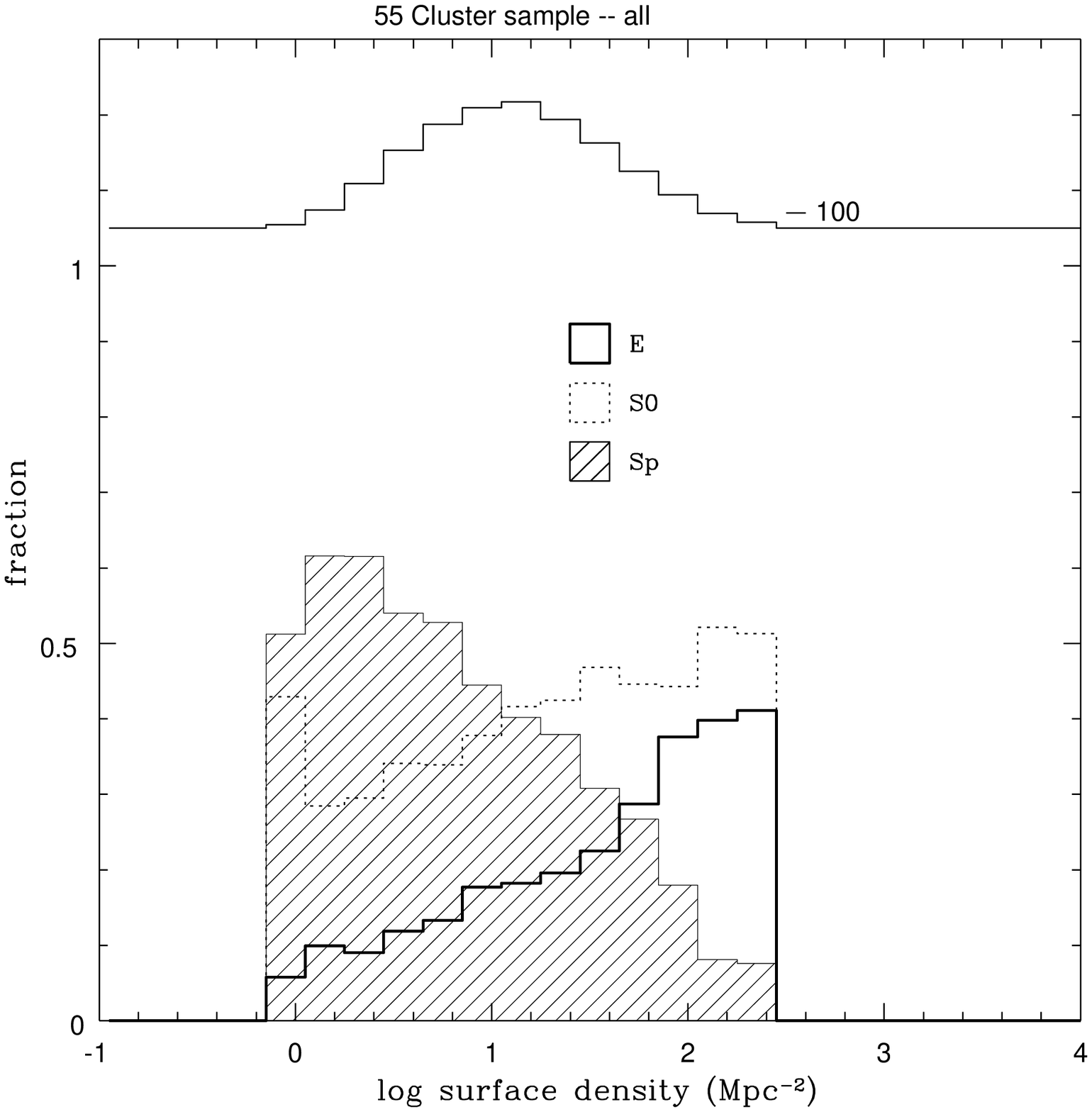,width=70mm}}
\caption{\it Evolution in the morphological fraction-projected surface
density relation in the central 1.2 $h^{-1}$ Mpc regions of rich
clusters from the analysis of Dressler et al (1997). (Top) 10 clusters
with 0.37$<z<$0.57 from the Morphs HST dataset, (Bottom) 55 local clusters
reanalysed from the Dressler (1980) sample. The histogram in the
top boxes show the number of galaxies in each bin.}
\end{figure}

Can the physical origin of the Butcher-Oemler effect really be so
simple? Sandage \& Visvanathan (1978) and Bower et al (1992)
demonstrated the significance of the small scatter in the rest-frame UV
colour-magnitude relation.  Using the Morphs data, Ellis et al (1997)
find a remarkably small scatter ($<$0.07 mag in rest-frame U-V) in the
colours of the elliptical population, even across different clusters,
at z$\simeq$0.5 (see Stanford et al 1997 for an extension over a wider
redshift range).  The bulk of this population must be
passively-evolving systems whose stars formed at an early time ($z>$3)
in broad agreement with earlier ground-based work (Aragon-Salamanca et
al 1993). Similar constraints are now emerging from fundamental plane
(Mg - $\sigma$ and $r_e$ - $\sigma$) relations pursued to high redshift
(van Dokkum \& Franx 1996, Bender et al 1997, Kelson et al 1997).

However, although {\it some} cluster ellipticals are undoubtedly old, 
it is by no means clear that {\it all} of the present day population 
formed in this way. Ellipticals could be continuously formed e.g. via
mergers as clusters assemble, in a way that guarantees that at any epoch
only the most quiescent, morphologically regular examples show minimal
UV scatter. Only number density arguments can circumvent this bias
and cluster samples are poorly-suited to such analyses. Indeed, 
Kauffmann (1995) has suggested that the richest systems selected
at high $z$ are likely to be observed more immediately after their
assembly than is the case locally and this would complicate any
attempt to construct volume-limited samples for tracking the various
populations.

Moreover, not {\it all} the present-day S0s can be transformed spirals
either, as those few found at high redshift share the tight scatter of
the ellipticals (Ellis et al 1997). And of course, red S0s are found in
the local field which, presumably, cannot have been affected by those
processes we claim are restricted to dense environments. To verify the
simple picture above, one would ideally like to find examples of
objects in transition. Smail et al (1997b) have penetrated further down
the colour-luminosity relation in intermediate redshift clusters and
found a marked increase in the UV scatter for otherwise passively-evolving
systems. The dichotomy could be resolved following earlier conjectures 
that present-day S0s are of two types (van den Bergh 1990): luminous 
examples formed at birth and less luminous examples which may have formed 
from destroyed spirals. In this case, one would expect to locate strong
luminosity trends in the recent evolution of disk galaxies. As may also
be the case for the ellipticals, distinctive though a Hubble class may 
appear to be, not only may it be a transient phenomenon but there may 
also be a number of routes to each present-day taxonomic class.

What then are the {\it physical} processes which led to the remarkably
recent postulated transformation of spirals into S0s? A number of 
hypotheses have been advanced including dynamical friction in the 
cluster potential, gas stripping of infalling field galaxies by a 
dense intracluster medium and galaxy-galaxy merging induced perhaps 
by the hierarchical growth of clusters (see Oemler 1992 for a summary 
of the pre-HST data in the context of these theories). Numerical 
simulations of some of these effects have been performed by Mihos (1995) 
and by Moore et al (1996). Semi-analytical calculations have been
performed by Balland et al (1997).

Significant advances are now being made by linking the HST imaging data
with spectroscopic diagnostics of recent star formation (Couch et al
1994, Dressler et al 1994, Barger et al 1996, Couch et al 1997). A
surprising fraction of the spectrally active systems (strong [O II]
emission, deep Balmer absorption lines and blue continua) show signs of
morphological disturbances; many are suggestive of {\it merging
systems}, virtually always involving a disk galaxy (Figure 3). In
contrast, the spectrally evolved ``post-burst" systems (no [O II], deep
Balmer absorption lines and red continua) are primarily regular Es or
S0s (Figure 3). One puzzle introduced by Barger et al in attempting to
define a single cycle of activity (driven by merging or strong
dynamical friction in the core) relates to the {\it luminosity
distributions} of these two classes. If the latter population follows
the former, a substantial fading in luminosity would be expected; in
fact the $K$-band luminosity distribution of the post-burst population
is comparable to that of the active group.

It seems unlikely, given the sample size now available, that the
absence of luminous blue precursors of the red post-burst galaxies
(Figure 3) is a selection effect arising from the short period when
they would be visible.  A more sensible explanation is that a large
proportion of the luminous post-burst systems had their star formation
{\it truncated} about 1-2 Gyr earlier, i.e. no burst occurred. Those
bursting galaxies that {\it are} observed presumably fade into systems
below the current survey limits ($M_V>$-20 + 5 log $h$) and thus their
fate is not yet clear. The existence of at least {\it two distinct
processes} contributing to the transformation occurring in the rich
cluster environment should not be that surprising.

Abraham et al (1996b) propose that truncated star formation initiated
for an infalling population of field galaxies could be an important
producer of S0s; in support they observe a strong radial gradient in
the mean Balmer line strength over 0.4 - 3 $h^{-1}$ Mpc in the
well-studied cluster Abell 2390.  Similar spectral signatures have been
seen in the outskirts of the Coma cluster (Caldwell et al 1993) and
other nearby groups (Caldwell \& Rose 1997) are interpreted in the
context of infalling groups whose members suffer some event after a
first passage through the intercluster medium in the core. One major
puzzle remains with truncated models which bypass the blue burst phase
however, namely that the depth of the observed Balmer absorption lines
cannot easily be reproduced (Poggianti \& Barbaro 1996).
  
\begin{figure}
\psdraft
\psfull
\caption{\it WFPC-2 F814W images of galaxies with
intermediate dispersion spectra from the cluster study of Couch
et al (1997). Starburst galaxies (top 3) and blue post-starburst galaxies
(next top 3) indicate merging is a likely cause for the enhanced activity.
Red galaxies with strong H$\delta$ absorption (bottom 6) indicate an 
advanced stage of decline after recent star formation. Most of these 
are too luminous to be likely descendents of the starburst galaxies.}

\end{figure}

To summarise, the above studies have been invaluable in demonstrating
convincingly that environmental processes {\it do} influence galaxy
evolution and that the morphology-density relation was produced
remarkably recently.  The results highlight the transient nature of
morphology and consequent uncertainties associated with assuming fixed
population densities over a range in look-back time. The combination of
spectral diagnostics and HST imaging has been particularly effective.
It seems that more than one process may be responsible for the demise of the
cluster spirals seen in the HST image. In the cluster core, merging and
dynamical friction may excite the abundant gas-rich disks into fresh
starbursts and these may fade into lower luminosity spheroidals.
However, some S0s, perhaps the most luminous examples in the cluster
cores, appear to be genuinely old along with the bulk of the
ellipticals. A continuous infall of field galaxies may provide an
additional source of material for transformation.  The overall goal
is now to reproduce the luminosity functions of the various
morphological types in well-defined samples at various redshifts
according to the processes responsible.

\section{The Evolution of Field Galaxies to $z\simeq$1}

The lessons learnt in the cluster studies concerning the transient
nature of a morphological type apply even more so in the lower density
`field' environment where the changes with look-back time are more
dramatic (Ellis 1997). Following the theme of synergy between ground
and space, there are two results, namely (i) the synthesis of HST
imaging with two major redshift samples, the Canada-France (CFRS, Lilly
et al 1995) and LDSS (Ellis et al 1996) surveys, and (ii) the
integration of these results with deeper work much of which has been
inspired by the Hubble Deep Field (HDF, Williams et al 1996).

The CFRS and LDSS redshift surveys reveal significant changes in the
luminosity function (LF) have occurred since $z\simeq$1 that are likely
to be luminosity-and colour-dependent. The trends are strongest when
attention is confined to the star-forming component (Ellis 1997). Galaxy counts
subdivided by HST morphology from the Medium Deep Survey (MDS,
Glazebrook et al 1996, Driver et al 1996) likewise indicate dominant
trends for a sub-population of galaxies with irregular morphology. Do
these results mean the bulk of the recent field evolution is restricted
to one sub-luminous class of galaxy?

The availability of HST morphology and spectroscopic redshifts
for a large homogeneous sample has been slow to emerge, because of the
mismatch in field of view between WFPC-2 and efficient ground-based
multiobject spectrographs. By mosaicing WFPC-2 images of ground-based
redshift survey fields, the CFRS and LDSS teams have together
constructed a sample of over 300 galaxies with $I<$22 and $B<$24
spanning the redshift range 0$<z<$1.2. A high proportion of the sample
fall into the category of galaxies with irregular morphology, with good
agreement between those so defined visually and via the automated
measurement of assymetry and concentration. With the benefit of
redshifts, the contribution of these sources to the overall redshift
trends can, for the first time, be examined (Brinchmann et al 1997).

\begin{figure}
\centerline{\psfig{file=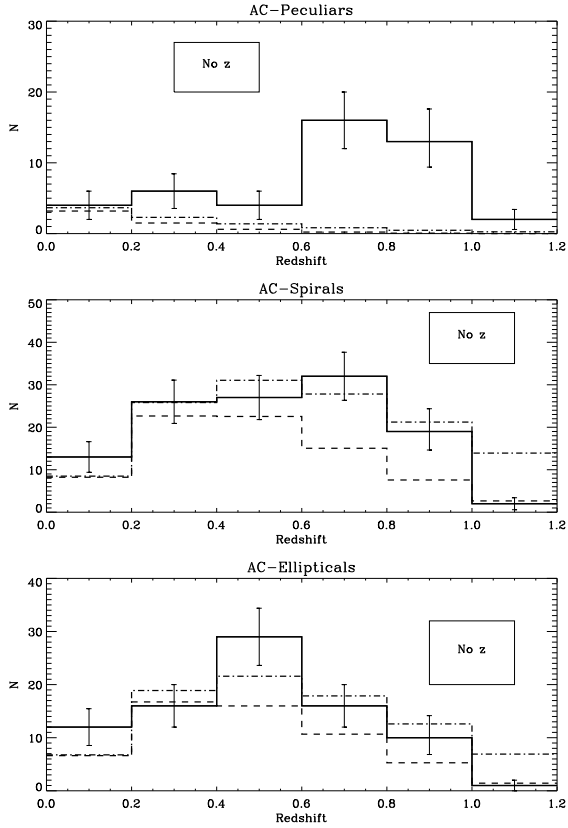,width=80mm}}
\caption{\it The redshift distribution of galaxies in the CFRS/LDSS
HST imaging survey divided according to a morphological criterion
based on concentration and asymmetry (after Brinchmann et al 1997).
AC-P, AC-S, AC-E refer respectively to irregular/peculiars, spirals
\& spirals. Model predictions assume no evolution (dashed) and 
mild evolution (dash-dotted, corresponding to 1 mag. of
luminosity evolution in rest-frame $B_{AB}$ at $z$=1). }
\end{figure}

The new survey indicates a dramatic increase with redshift in the
fraction of irregular sources (Figure 4), significantly more than can
be accounted for by late-type spirals misclassified as irregulars
because of redshift-dependent effects (Brinchmann et al 1997). By
contrast only modest luminosity evolution is necessary to account
for the redshift distribution of the regular spirals and ellipticals. 
Most importantly, the comoving volume density at $z\simeq$0.8 is 
comparable to local estimates. Most of the rise with redshift in 
the blue luminosity density appears to originate in the irregular/peculiar 
population although spirals still dominate the overall
flux at $z\simeq$1 (Figure 5).

\begin{figure}
\centerline{\psfig{file=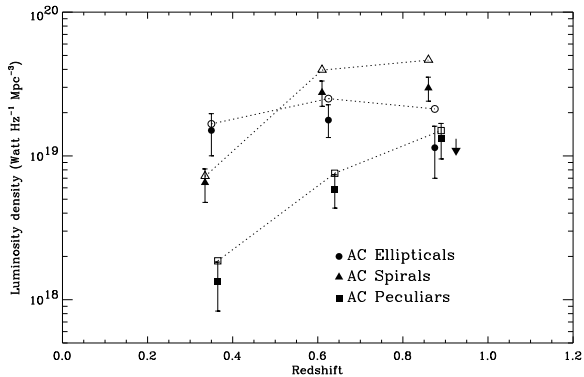,width=80mm}}
\caption{\it The $B_{AB}$ luminosity density of galaxies detected in
the CFRS/LDSS HST imaging survey as a function of redshift and class.
The downward arrow indicates the degree to which redshift-dependent
effects cause a spurious migration of true spirals into apparent
irregular/peculiars. Open symbols refer to estimated integrated values
extrapolating the luminosity function below the detection limits (after
Brinchmann et al 1997).}
\end{figure}

Do the distant irregulars represent a truly independent population
of rapidly-evolving systems or are they normal systems rendered
irregular by increased star formation? One indication that the irregulars
are not ``agitated spirals" is that their marked decline in number
as redshift decreases is not accompanied by any obvious increase in 
the number density of regular systems. It turns out this is not a 
particularly convincing argument, however, since the star forming 
irregulars would presumably fade as they turned into well-behaved 
galaxies and many of these may lie below the CFRS/LDSS magnitude limits.  

A more convincing measure of the slowly-evolving population of disk
galaxies can be derived from structural parameters such as bulge/disk
ratios, scale lengths and central surface brightnesses following
profile fitting techniques developed by Schade et al (1996). A disk
size function, $\Phi(\alpha^{-1}$ kpc) Mpc$^{-3}$, has been constructed
by Lilly et al (1997) taking account of flux and surface brightness
limits using an analog of the $V/V_{max}$ test. This size function is
only moderately sensitive to the cosmological world model and indicates
that large spirals ($\alpha^{-1}>$2$h^{-1}$ kpc) have more or less
the present volume density at $z\simeq$0.8. Surface photometry suggests
only modest luminosity evolution ($<$0.5 mag) and limited size growth ($<$20\%)
over the redshift range of the survey (0.3$<z<$1).
    
What then is the physical origin of the star forming irregulars which
dominate the evolutionary changes seen to $z\simeq$1? Broadhurst et al
(1988) proposed, on the basis of spectroscopic arguments similar to
those discussed in $\S$2, that sub-luminous galaxies are being rendered
visible by short-lived bursts of star formation. Both infrared
luminosities (Cowie et al 1996) and emission line widths (Guzman et al
1996) indicate the rapidly-evolving objects have low masses. Depending
on the physical origin of this enhanced star formation and the nature
of the underlying stellar population, it is easy to see how an
irregular system would be produced for a limited period. Alternatively,
some appear to be truly compact and similar to extragalactic H II
regions (Guzman et al 1997) whilst others appear to be merging (LeFevre
et al 1997).

The Broadhurst et al burst model unfortunately implies the presence of an even 
greater abundance of low mass systems in a quiescent state which serve 
as a reservoir for this activity. The outstanding problem is to find 
examples in the quiescent phase and ultimately the present day faded 
remnants of this population (amusingly, this is the opposite of the problem we 
encountered in $\S$2). With a normal initial mass function, it is difficult 
to exceed a fading of more than 3 magnitudes in rest-frame $B$ since
z$\simeq$0.5 (Phillipps \& Driver 1995, Babul \& Ferguson 1996) so the 
absence of a dominant component of {\it red} systems with 
$M_B\simeq$-16 + 5 log $h$ remains a puzzle. 

We might make progress towards resolving this long-standing riddle
along the following lines.  Foremost, it is reasonable to support that
the rapidly evolving component could be the product of {\it several}
processes that might befall an abundant population of gas-rich dwarfs
expected at early times. The classification of the star forming
galaxies outside the range of the regular Hubble sequence has perhaps
mistakenly led to a search for a single physical process whereas in
fact, one of {\it several} mechanisms that leads to an increase in star
formation rate can produce these systems. For each mechanism
(agitation, fading, disruption, merging, misclassification), a
different ``remnant" results and meeting the constraints on each would
not be so demanding. This is particularly so when it is realised that
our  knowledge of the present day galaxy population is fairly
inadequate at $M_B>$-16, especially in absolute volume densities (Ellis
1997).

In summary, the combination of HST and ground-based redshift surveys
has shown a surprising contrast between the modest evolution found
for the regular disk galaxies and the large increase in number density
for irregular systems. Only a small fraction of this irregular population
can be explained through mistaken classification problems arising from
surface brightness and redshift effects. Whilst it is tempting to
search for a single mechanism to explain this dominant activity in the
evolution of the luminosity density to $z\simeq$1, possibly the
phenomenon is the manefestation of several distinct processes whose
common theme is simply increased star formation (Figure 5). 
 
\section{Beyond $z\simeq$1}

With the systematic study of the $z<$1 population well underway, it is
appropriate to now consider the best strategies for surveying the
population beyond a redshift of 1. The `redshift one barrier' was
traditionally broken only by locating spectacularly luminous active
galaxies (Lilly 1988, Chambers et al 1990, Rowan-Robinson et al 1992)
or gravitationally-magnified sources (Ebbels et al 1996, Franx et al
1997). More recently, through Lyman-limit imaging and Keck spectroscopy
a population of luminous star forming galaxies has been systematically
explored in the redshift window 2.3$<z<$4.0 (for a recent review see
Pettini et al 1997). This much-heralded observational breakthrough has
been accompanied by considerable investigation of the robustness of the
Lyman-limit selection criteria (Madau et al 1996), the physical
significance of the properties of the galaxies so found (Steidel et al
1996,1997) and the understanding of the results in the context of
hierarchical pictures (Baugh et al 1997).

A surprising result from the Lyman-limit galaxy surveys, but one that
was, in fact, tentatively suggested from earlier results (Guhathakurta
et al 1990) is that the inferred star formation rate per unit comoving
volume at $z\simeq$3 is considerably lower than that at $z\simeq$1. At
the current moment, more is known about the nature of the visible
galaxy population at $z\simeq$2-3 than immediately beyond $z$=1. This
curious state of affairs is purely a by-product of the technique used
to locate the high $z$ sources which cannot yet be implemented at
redshifts below $z\simeq$2.3. Despite the rapid progress, ultimately we
will still need to extend the traditional redshift surveys beyond $z$=1
(see $\S$5).

The rising star-formation density to $z\simeq$1 and the inferred
decline thereafter can be tracked with some uncertainties by inferring
redshifts from {\it multiband photometry}. In this way, Connolly et al
(1997) analyse the HDF dataset to delineate a history of the
volume-averaged star formation rate (SFR) whose peak occurs at a
surprisingly low redshift. However, in the all-important region
1$<z<$2, the reliability of the photometric redshift technique is 
largely unchecked. Moreover, whereas the validity of the technique is 
often justified by comparisons with spectroscopic redshifts to $I\simeq$24 
(Lanzetta et al 1997, Hogg et al 1997), in practice the 
{\it science conclusions} are based on an application of the method 
to much fainter data. A cross-comparison between two independent
applications to the HDF dataset to $I<$26 revealed disappointing
agreement (Figure 6). Furthermore, a fundamental assumption of
the method as commonly used is that the spectral energy distribution 
of a high redshift galaxy evolves in such a way that it always
somehow lies along the locus of present-day equivalents. Given the
number of puzzling results discussed for $z<$1 galaxies ($\S$2,3)
this seems unjustified.

\begin{figure}
\centerline{\psfig{file=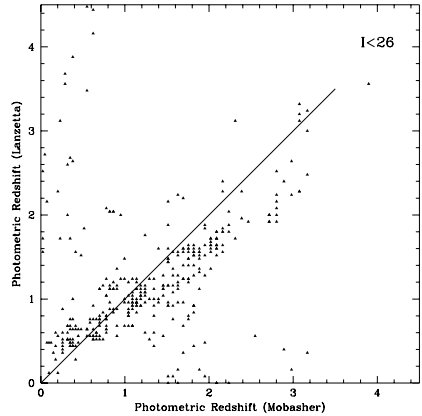,width=60mm}}
\caption{\it A comparison of the published photometric redshifts from
Lanzetta et al (1996) and Mobasher et al (1996) for the Hubble Deep
Field to $I$=26, i.e. well beyond the limits of Keck spectroscopy
(Ellis 1997).}
\end{figure}

The photometric redshift technique would be useful if it could be proven 
to yield reliable redshifts at the $\Delta\,z\simeq\pm$0.3 level to, say,
$I$=25. This is a reasonable goal to verify given the spectroscopic 
capabilities of 8-10m telescopes. Currently, however, there is little
evidence that beyond $z$=1 where useful information is sought. Indeed,
there is not yet convincing evidence that the precision of the multi-band
fitting method is any better than that indicated by the simple location of 
the Lyman limit. Nonetheless, even this crude precision is sufficient 
to support the overall picture of a recent peak of star formation.

In interpreting the star formation histories quantitatively, several
caveats apply. Firstly, the SFR is based only on the {\it detected}
emission from galaxies at optical and uv wavelengths. Dust will not
only lead to underestimates of the SFR but also obscured sources will
re-radiate at wavelengths where there is currently very little data;
this needs to be accounted for in the global calculations (Fall et al
1996, Madau et al 1996). Secondly, different techniques have been
applied to infer the SFR in different redshift regimes, and flux
limited samples have been corrected to integrated values by making
assumptions about the undetected population. Madau (1997) discusses the
limitations in detail. With the exception of dust, it is difficult to 
see how any of these caveats could seriously distort the qualitative 
picture.

Recent discussion concerning dust has centred on two themes. Firstly,
the spectra of the Lyman-limit galaxies show features supportive of
high star formation rates yet, puzzlingly, the slopes of their UV
continua do not match that expected for young stellar populations
(Pettini et al 1997, Calzetti 1997). If this discrepancy is due to
dust, it would lead to a fairly modest upward correction in the star
formation density. More serious is the objection, from studies at far
infrared and sub-mm wavelengths (Mann et al 1997, Rowan-Robinson et al
1997, Smail et al 1997c), that even a small number of sources
indicative of very high SFRs found at these wavelengths cannot readily
be reconciled with the SF history derived from primarily optical data.
At the time of writing, the statistics justifying this view are
hopelessly inadequate and a key point here is to understand how
dramatic modifications to the `hierarchically approved' SF history
could be reconciled with the low metallicity of the Lyman-limit
galaxies (Lowenthal et al 1997), the Lyman alpha forest (Songaila \&
Cowie 1996) and current limits on the far-IR background (Puget et al
1996).

\section{The Future}

The picture emerging is only empirical and the greatest progress made
has been concerned with the {\it integrated} properties of the galaxy
population. To make progress in understanding the physical basis of the
evolutionary trends - for example to test whether the mass assembly
rate in stars is consistent with hierarchical theories - we must begin
to break the samples into subclasses located by colour, line strength
and HST morphology. The above discussion indicates the way each of
these observables may be transient and hence not conserved in
populations based on normal observational selection.  Accordingly, 
we will need to be much more imaginative in securing and intepreting 
data that will rise to this challenge.

I suggest there are two logical next steps. The first is to systematically 
explore the supposed ``peak" of SF activity at 1$<z<$2 spectroscopically.
Paradoxically this region is relatively unexplored notwithstanding 
the progress made via Lyman limit surveys at higher redshift. 
The only reliable technique to extend the luminosity function studies 
begun so successfully at $z<$1 requires $K$-selected spectroscopy 
with near-infrared instruments sensitive to narrow emission lines.
As the IR background is dominated in the $J$ and $H$ windows
by OH emission, its suppression via hardware or software methods
is essential. Given the preponderance of blue star-forming systems
at these limits, emission line-based surveys should be highly
successful (Figure 7).

\begin{figure}
\centerline{\psfig{file=iau_fig7a.ps,width=60mm,angle=270}}
\centerline{\psfig{file=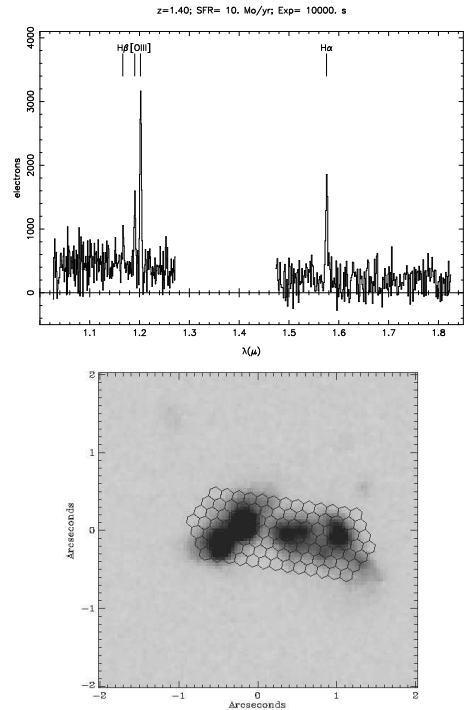,width=60mm}}
\caption{\it (Top) Simulated infrared spectrum for a star forming galaxy
at $z$=1.4 as it would be observed with a 4-m telescope using the 
Cambridge OH-suppression infrared spectrograph (COHSI, Pich\'e et al 1997) . 
Such instrumentation will allow an extension of well-proven spectroscopic
techniques into the important redshift range 1$<z<$3. (Bottom) COHSI's 
integral field unit overlaid on the HST image of a $z$=1.355 HDF galaxy
illustrating the possibility of resolved spectroscopic and dynamical data. 
Each lenslet samples a field of diameter 0.15 arcsec ($\simeq$0.6 $h^{-1}$ kpc).}
\end{figure}

The second step is to exploit the {\it resolved} data on high $z$ sources
made available from HST in order to commence linking the trends
observed to dynamical data. Line widths (Guzman et al 1996)
and rotation curves (Vogt et al 1996) represent an important step
forward, but 2-D velocity fields for selected irregular galaxies, such
as those in the HDF, should also be attempted using the new generation
of integral field spectroscopic units and linked with multicolour 
analyses (Abraham et al 1997, Figure 7). The challenge of interpreting such
data will be considerable but it is now necessary to move the subject 
from statistical descriptions of an empirical nature into ones more 
closely related to galactic structure, dynamics and the astrophysical
origin of star formation on galactic scales. 

\section{Acknowledgements}

I acknowledge numerous discussions with my colleagues at Cambridge and
collaborators on the Morphs, redshift survey, HST and infrared
instrumentation programmes. I thank the IAU organisers and colleagues
at NAOJ for generous travel funds that enabled me to participate 
in this symposium. I also thank the Carnegie Observatories in Pasadena 
for their hospitality and support during a period of sabbatical leave 
where this article was completed.

\end{document}